\documentclass[journal,final,twoside]{IEEEtran}

\usepackage[noadjust]{cite}
\usepackage[dvipsnames]{xcolor}
\usepackage[normalem]{ulem}
\usepackage{cuted}
\usepackage[hidelinks]{hyperref}
\usepackage{dblfloatfix} 
\ifCLASSINFOpdf
  \usepackage{graphicx}
  \graphicspath{{figures/}{../jpeg/}}
  \DeclareGraphicsExtensions{.eps,.pdf,.jpeg,.png}
\else
\fi

\ifCLASSOPTIONcompsoc
    \usepackage[caption=false, font=normalsize, labelfont=sf, textfont=sf]{subfig}
\else
    \usepackage[caption=false, font=footnotesize]{subfig}
\fi

\usepackage{amsmath,amssymb,bm}
\usepackage{xfrac}
\usepackage{upgreek}
\usepackage{siunitx}
\DeclareSIUnit{\belmilliwatt}{Bm}
\DeclareSIUnit{\belwatt}{BW}
\DeclareSIUnit{\dBm}{\deci\belmilliwatt}
\DeclareSIUnit{\dBW}{\deci\belwatt}

\usepackage{stackengine}

\usepackage{algorithm}
\usepackage{algpseudocode}

\usepackage[acronym]{glossaries}
\newacronym{ber}{BER}{bit error rate}
\newacronym{cfo}{CFO}{carrier frequency offset}
\newacronym{df-kic}{DF-KIC}{decision feedback-aided known-interference cancellation}
\newacronym{emse}{EMSE}{excess mean-squared error}
\newacronym{evm}{EVM}{error vector magnitude}
\newacronym{fd}{FD}{full-duplex}
\newacronym{fir}{FIR}{finite impulse response}
\newacronym{fo-lms}{FO-LMS}{frequency offsets-compensated least mean squares}
\newacronym{ki}{KI}{known interference}
\newacronym{kic}{KIC}{known-interference cancellation}
\newacronym{lms}{LMS}{least mean squares}
\newacronym{mse}{MSE}{mean-squared error}
\newacronym{nlms}{NLMS}{normalized least mean squares}
\newacronym{ocxo}{OCXO}{oven-controlled crystal oscillator}
\newacronym{ofdm}{OFDM}{orthogonal frequency division multiplexing}
\newacronym{rls}{RLS}{recursive least squares}
\newacronym{sdr}{SDR}{software-defined radio}
\newacronym{ser}{SER}{symbol error rate}
\newacronym{sfo}{SFO}{sampling frequency offset}
\newacronym{sgd}{SGD}{stochastic gradient descent}
\newacronym{si}{SI}{signal of interest}
\newacronym{sinr}{SINR}{signal-to-interference-and-noise ratio}
\newacronym{snr}{SNR}{signal-to-noise ratio}
\newacronym{src}{SRC}{sampling rate conversion}
\newacronym{tcxo}{TCXO}{temperature-compensated crystal oscillator}
\newacronym{vss-fo-lms}{VSS-FO-LMS}{variable step sizes frequency offsets-compensated least mean squares}
\newacronym{vss-nlms}{VSS-NLMS}{variable step size normalized least mean squares}

\begin{document}

\title{Decision Feedback-Aided\\Known-Interference Cancellation}

\author{Karel~P{\"a}rlin,
        Aaron~Byman,
        Tommi~Meril{\"a}inen, and
        Taneli~Riihonen
\thanks{This work has been submitted to the IEEE for possible publication. Copyright may be transferred without notice, after which this version may no longer be accessible.}
\thanks{This research work was supported by the Research Council of Finland and the Finnish Research Impact Foundation.}
\thanks{K.~P{\"a}rlin and T.~Riihonen are with Tampere University, Faculty of Information Technology and Communication Sciences, Korkeakoulunkatu~1, 33720 Tampere, Finland (e-mail: karel.parlin@tuni.fi; taneli.riihonen@tuni.fi).}
\thanks{A.~Byman and T.~Meril{\"a}inen are with Bittium Wireless, Ritaharjuntie~1, 90590 Oulu, Finland (e-mail: aaron.byman@bittium.com; tommi.merilainen@bittium.com).}
}

\markboth{}%
{P{\"a}rlin \MakeLowercase{\textit{et al.}}: Decision Feedback-Aided Known-Interference Cancellation}

\maketitle

\begin{abstract}
\Gls{kic} in combination with cooperative jamming can be used to provide covertness and security to wireless communications at the physical layer. However, since the \gls{si} of a wireless communication system acts as estimation noise, i.e., interference, to \gls{kic}, the \gls{si} limits the extent to which the \gls{ki} can be canceled and that in turn limits the throughput of the wireless communication system that is being hidden or secured. In this letter, we analyze a \gls{df-kic} structure in which both the \gls{ki} and \gls{si} are canceled iteratively and successively. Measurement results demonstrate that introducing decision feedback to \gls{kic} improves its \gls{ki} cancellation capability and hence increases the wireless communication system's useful throughput, albeit at the expense of a higher computational load.
\end{abstract}


\IEEEpeerreviewmaketitle

\section{Introduction}

\IEEEPARstart{W}{ireless} communications are inherently susceptible to adversarial exploits, such as detection and eavesdropping, on the physical layer level. Cooperative jamming and receiver-side \gls{kic} are one way to fortify wireless communications against this weakness~\cite{hamamreh2018classifications, parlin2023known, guo2019testbed}. The key challenge in cooperative jamming is interference management and its suppression~\cite{atallah2015survey}, the latter of which requires at the very least precise time--frequency synchronization and channel estimation \cite{guo2018effect, guo2020comprehensive, guo2021analysis, parlin2023estimating} but can also benefit from estimating other hardware impairments~\cite{he2021performance, parlin2025high, guo2025cooperative}. Numerous works on \gls{kic} have demonstrated the feasibility of estimating and compensating for these aspects relying on different sets of hardware operating in both laboratory and outdoor conditions~\cite{guo2019testbed, guo2021analysis, parlin2023known, parlin2024wideband, parlin2025high, parlin2025rayleigh}. However, as the \gls{ki} propagation path modeling becomes sufficiently accurate, in certain cases the limiting factor to \gls{kic} becomes instead the \gls{si}, since the \gls{si} ultimately acts as estimation noise to the adaptive \gls{ki} cancellation. This will prevent the post-\gls{kic} channel capacity from reaching that what would be achieved without cooperative jamming.

\newpage
One way to approach the post-\gls{kic} residual \gls{ki}, and to approach interference from a wireless communications perspective in general, is to adapt the \gls{si} transmission rate to match what the interference-included channel capacity facilitates~\cite{love2008overview, hanawal2016joint}. However, this will of course decrease the communication system's throughput~\cite{qiu1999performance}. The alternative way is to try to reduce the impact that the different received signals have on one another in their respective processing stages. Since the output of a receiver is an estimate of the transmitted \gls{si} symbols, that estimate can be typically used to reconstruct a replica of the transmitted \gls{si} waveform, which can then be suppressed in the received signal like the \gls{ki}~\cite{milstein2002interference}. If any of the initial estimates of the transmitted \gls{si} symbols are erroneous, which assumedly is the case, since otherwise improving \gls{kic} would not be necessary, such decision feedback-aided filtering has of course the possibility of error propagation. 

Still, in many cases the methods with decision feedback outperform those without it, e.g., in narrowband interference cancellation~\cite{ketchum1984decision, takawira1987error, shah1988adaptive, dukic1990performance, laster1997interference}, in channel equalization~\cite{george1971adaptive, belfiore1979decision, kaleh1995channel}, and in multi-user interference cancellation~\cite{duel1995multiuser, wang1999iterative, andrews2005interference}. In this letter, we therefore propose a \gls{df-kic} method that iteratively cancels both the received \gls{ki} and \gls{si}, whereas the transmitted digital waveform of the former is known completely \textit{a priori} and the latter is of a known structure, e.g., OFDM, but contains unknown data. We demonstrate the method's performance in comparison to that without decision feedback based on laboratory measurements carried out with commercial off-the-shelf \glspl{sdr} and using \glspl{si} with different \gls{sinr} requirements.

\section{Known-Interference Cancellation}
\label{sec:kic}

\subsection{System Model and Base Known-Interference Cancellation}
\label{ssec:system_model}

The \gls{ki} and \gls{si} transmitters broadcast signals $x(n)$ and $s(n)$, respectively, the former of which is known \textit{a priori} to only authorized receivers and the latter of which is unknown but of interest to both authorized and unauthorized receivers. Then, the superposed signal at any receiver becomes
\begin{equation}
    d(n) = \mathbf{w}_n^H \mathbf{y}_n e^{j\phi^\mathrm{x}(n)} + \mathbf{h}_n^H \mathbf{z}_n e^{j\phi^\mathrm{s}(n)} + v(n),
\end{equation}
where $\mathbf{w}_n$ and $\mathbf{h}_n$ are the channel impulse responses from the \gls{ki} and \gls{si} transmitters to the receiver, respectively, $\left\{ \cdot \right\}^H$ denotes conjugate transpose, $v(n)$ is measurement noise with variance $\sigma^2_\mathrm{v}$, $\mathbf{y}_n$ and $\mathbf{z}_n$ account for $x(n)$ and $s(n)$ with time-varying sampling frequency offsets $\eta^\mathrm{x}(n)$ and $\eta^\mathrm{s}(n)$ according in~\cite[Eq.~(2)]{parlin2023estimating}, and the multiplicative terms $e^{j\phi^\mathrm{x}(n)} = e^{j\sum^n_{i=1}{\epsilon^\mathrm{x}(i)}}$ and $e^{j\phi^\mathrm{s}(n)} = e^{j\sum^n_{i=1}{\epsilon^\mathrm{s}(i)}}$ account for the carrier frequency offsets and phase noise.

Not knowing $x(n)$, any unauthorized receiver is left with the superposition of the signals. Authorized receivers, however, can suppress $x(n)$ from the received signal so that
\begin{equation}\label{eq:mse}
    e(n) = d(n) - \hat{\mathbf{w}}^H_{n} \hat{\mathbf{y}}_n e^{j\hat{\phi}^\mathrm{x}(n)}
    \stackrel{\text{ideally}}{=} \mathbf{h}_n^H \mathbf{z}_n e^{j\phi^\mathrm{s}(n)} + v(n),
\end{equation}
where $\hat{\mathbf{w}}_{n}$, $\hat{\epsilon}^\mathrm{x}(n)$, and $\hat{\eta}^\mathrm{x}(n)$ are, respectively, accurate estimates of the channel impulse response, carrier frequency offset, and sampling frequency offset at iteration $n$, $e^{j\hat{\phi}^\mathrm{x}(n)} = e^{j\sum^n_{i=1}{\hat{\epsilon}^\mathrm{x}(i)}}$, and $\hat{\mathbf{y}}_n$ is the result of resampling $x(n)$ with $\hat{\eta}^\mathrm{x}(n)$. In practice, estimation inaccuracies mean that some residual \gls{ki} will typically remain in \eqref{eq:mse}. For the base, i.e., non-decision feedback-aided, \gls{kic} we herein use the \gls{vss-fo-lms} algorithm, which iteratively updates its estimates $\hat{\mathbf{w}}_{n}$, $\hat{\epsilon}^\mathrm{x}(n)$, and $\hat{\eta}^\mathrm{x}(n)$ by minimizing the error in \eqref{eq:mse} based on \gls{sgd} rules~\cite[Algorithm 1]{parlin2025vssfolms}.

\subsection{Known-Interference Cancellation with Decision Feedback}
\label{ssec:algorithm}

\begin{figure}[htpb]
	\centerline{\includegraphics{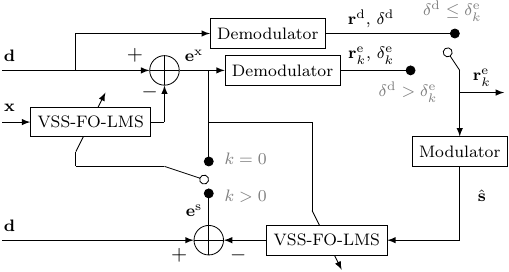}}
	\caption{Structure of \gls{df-kic} for successive cancellation of \gls{ki} and \gls{si}.}
	\label{fig:df-kic}
\end{figure}

The \gls{df-kic} steps are listed as Algorithm~\ref{alg:df-kic} and its structure is illustrated in Fig.~\ref{fig:df-kic}. The algorithm operates over an entire received \gls{si} frame and assumes that the start of the frame has been located outside of \gls{df-kic} (e.g., by relying on a known preamble and/or training symbols). The algorithm begins by initializing the estimated parameter sets $\zeta^\mathrm{x}$ and $\zeta^\mathrm{s}$ related to the \gls{ki} and \gls{si} channels, respectively, by setting the parameter estimates to zero. These sets will be passed across the decision feedback iterations to reduce the \gls{vss-fo-lms}'s convergence time, analogous to the concept of warm restarts in \gls{sgd} methods in general~\cite{loshchilov2017sgdr}.

The base \gls{kic} is then executed and both the original received signal $\mathbf{d}$ and the post-\gls{kic} error signal $\mathbf{e}^\mathrm{x}$ are demodulated. The demodulator is required to provide estimated data symbols $\mathbf{r}^\mathrm{d}$ and $\mathbf{r}^\mathrm{e}$ as well as received signal quality indicators $\delta^\mathrm{d}$ and $\delta^\mathrm{e}$. Herein, OFDM with different orders of QAM is used for modulation and nondata-aided \gls{evm}~\cite{9942923} is used as the quality indicator. Nondata-aided \gls{evm} is typically more optimistic than data-aided \gls{evm} but can still provide a useful estimate of the \gls{sinr}~\cite{mahmoud2009error, schmogrow2012error}. The quality indicator is relied on to assess which of the demodulated symbol streams, $\mathbf{r}^\mathrm{d}$ or $\mathbf{r}^\mathrm{e}$, is likely to contain less errors, whether that symbol stream is likely to be of sufficient quality for further processing, and whether successive decision feedback iterations improve the post-\gls{kic} \gls{sinr}.

If neither the received signal nor base \gls{kic} leads to a satisfactory quality indicator $\delta^\mathrm{t}$, the symbol stream with better quality is modulated to reconstruct an estimate $\hat{\mathbf{s}}$ of the transmitted \gls{si} waveform. That \gls{si} estimate is then matched to that received, using the base \gls{kic} and the error signal $\mathbf{e}^\mathrm{x}$, from which the \gls{ki} has been canceled, as the desired signal. The resulting estimate of the received \gls{si}, denoted as $\hat{\mathbf{d}}^\mathrm{s}$, is removed from the total received signal resulting in a \gls{si}-reduced received signal $\mathbf{e}^\mathrm{s}$. The base \gls{kic} is then used again to match the \gls{ki} to that received based on the \gls{si}-reduced received signal. The estimate of the received \gls{ki}, denoted $\hat{\mathbf{d}}^\mathrm{x}$, is removed from the total received signal and the resulting \gls{ki}-reduced received signal is demodulated, which concludes the decision feedback iteration.

If the stopping criterion $\delta^\mathrm{t}$ has not been met, if the signal quality indicator has improved, and if the maximum number of decision feedback iterations $K$ have not been reached, then the algorithm proceeds to the next iteration. Note that the \gls{vss-fo-lms} algorithm requires as inputs certain forgetting factors and the number of taps but these are herein assumed to be fixed and are omitted from the algorithm listing for brevity.

\begin{algorithm}
\caption{}\label{alg:df-kic}
\begin{algorithmic}[1]
\Procedure{DF-KIC}{$\mathbf{x}$, $\mathbf{d}$, $K$, $\delta^\mathrm{t}$}

\State $\zeta^\mathrm{x}(0) \gets \left\{\hat{\mathbf{w}}: \mathbf{0}, \hat{\epsilon}^\mathrm{x} : 0, \hat{\eta}^\mathrm{x} : 0\right\}$
\State $\zeta^\mathrm{s}(1) \gets \left\{\hat{\mathbf{h}}: \mathbf{0}, \hat{\epsilon}^\mathrm{s} : 0, \hat{\eta}^\mathrm{s} : 0\right\}$

\hskip\algorithmicindent
\State $\hat{\mathbf{d}}^\mathrm{x}$, $\zeta^\mathrm{x}(1)$ $\gets$ $\text{VSS-FO-LMS}\left(\mathbf{x},\ \mathbf{d},\ \zeta^\mathrm{x}(0)\right)$
\State $\mathbf{e}^\mathrm{x} \gets \mathbf{d} - \hat{\mathbf{d}}^\mathrm{x}$

\State $\mathbf{r}^\mathrm{d}$, $\delta^\mathrm{d} \gets \text{Demodulator}\left(\mathbf{d}\right)$
\State $\mathbf{r}_0^\mathrm{e}$, $\delta^\mathrm{e}(0) \gets \text{Demodulator}\left(\mathbf{e}^\mathrm{x}\right)$

\hskip\algorithmicindent
\If{$\delta^\mathrm{e}(0) > \delta^\mathrm{d}$}
    \State $\mathbf{r}_0^\mathrm{e} \gets \mathbf{r}^\mathrm{d}$
    \State $\delta^\mathrm{e}(0) \gets \delta^\mathrm{d}$
\EndIf
\If{$\delta^\mathrm{e}(0) < \delta^\mathrm{t}$}
    \State \textbf{end procedure}
\EndIf

\hskip\algorithmicindent
\For{$k \gets 1 \text{ to } K$}
\State $\hat{\mathbf{s}} \gets \text{Modulator}\left(\mathbf{r}_{k-1}^\mathrm{e}\right)$

\State $\hat{\mathbf{d}}^\mathrm{s}$, $\zeta^\mathrm{s}(k+1)$ $\gets$ $\text{VSS-FO-LMS}\left(\hat{\mathbf{s}},\ \mathbf{e}^\mathrm{x},\ \zeta^\mathrm{s}(k)\right)$
\State $\mathbf{e}^\mathrm{s} \gets \mathbf{d} - \hat{\mathbf{d}}^\mathrm{s} $

\State $\hat{\mathbf{d}}^\mathrm{x}$, $\zeta^\mathrm{x}(k+1)$ $\gets$ $\text{VSS-FO-LMS}\left(\mathbf{x},\ \mathbf{e}^\mathrm{s},\ \zeta^\mathrm{x}(k)\right)$
\State $\mathbf{e}^\mathrm{x} \gets \mathbf{d} - \hat{\mathbf{d}}^\mathrm{x}$
\State $\mathbf{r}_k^\mathrm{e}$, $\delta^\mathrm{e}(k) \gets \text{Demodulator}\left(\mathbf{e}^\mathrm{x}\right)$

\hskip\algorithmicindent
\If{$\delta^\mathrm{e}(k) < \delta^\mathrm{t}$}
    \State \textbf{end procedure}
\ElsIf{$\delta^\mathrm{e}(k) > \delta^\mathrm{e}(k-1)$}
    \State $\mathbf{r}_k^\mathrm{e} \gets \mathbf{r}_{k-1}^\mathrm{e}$
    \State \textbf{end procedure}
\EndIf

\EndFor
\EndProcedure
\end{algorithmic}
\end{algorithm}

\begin{figure}[htpb]
     \centering
     \subfloat[Diagram of the measurement setup\label{fig:measurement_model}]{
         \includegraphics[width=0.45\textwidth]{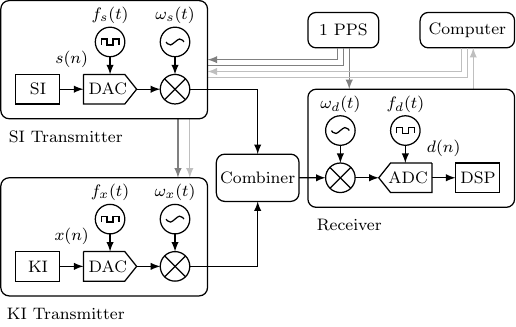}}
     \\
     \centering
     \subfloat[Photo of the measurement setup\label{fig:measurement_photo}]{
         \includegraphics[trim={0 0 0 0},clip,width=0.45\textwidth]{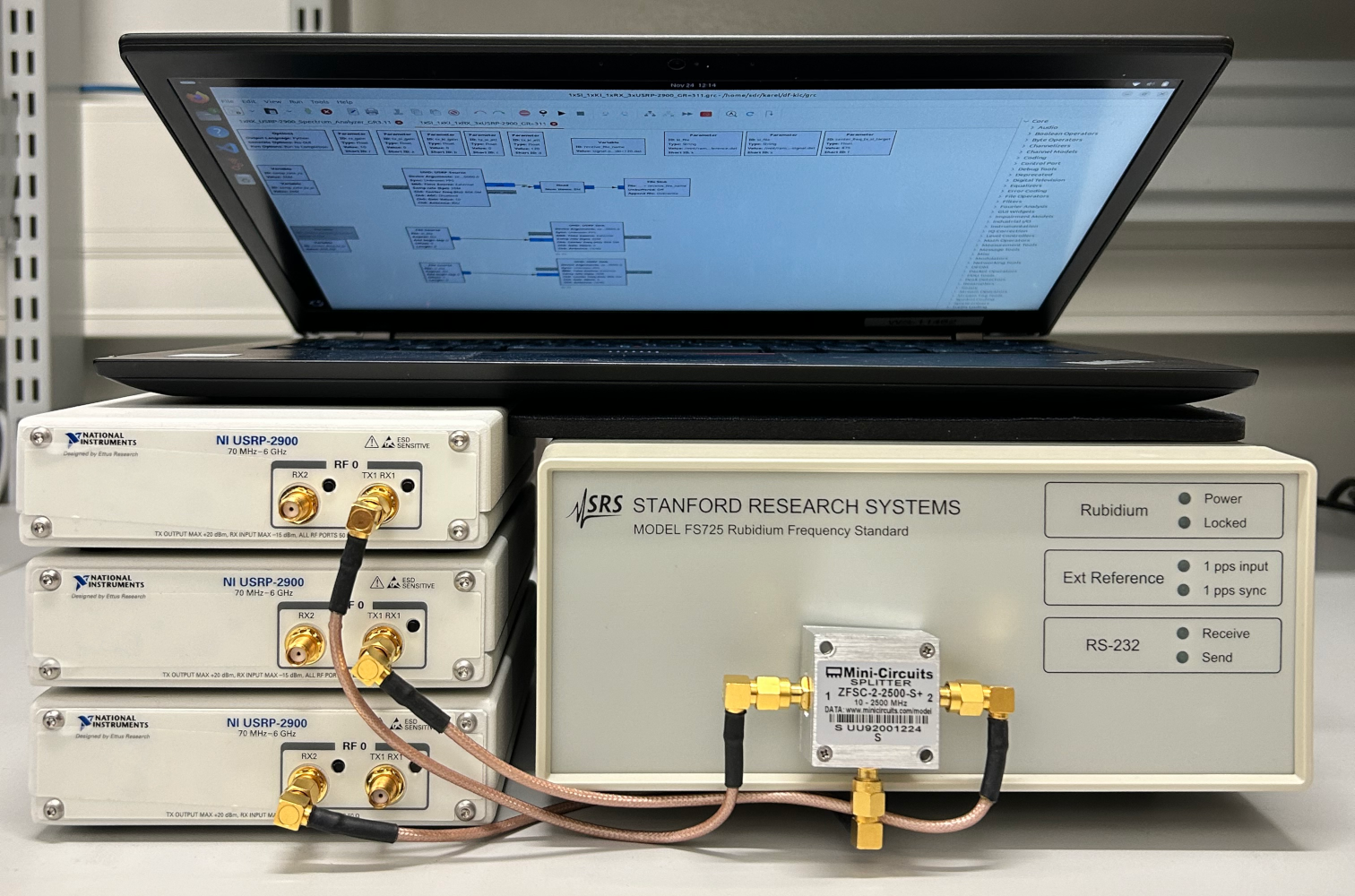}}
    \caption{Laboratory measurement setup with three USRP-2900 \glspl{sdr}.}
    \label{fig:measurement_setup}
\end{figure}

\section{Measurement Setup}
\label{sec:measurement_setup}

The impact of decision feedback iterations on \gls{kic} was studied based on measurements carried out using the laboratory setup illustrated in Fig.~\ref{fig:measurement_setup} and described below. The transmitters and the receiver were implemented using USRP-2900 \glspl{sdr}, which were connected over cables. A digital intermediate frequency was used to remove the impact of LO leakages and IQ imbalances. Effectively, the \glspl{sdr} were configured to \SI{875}{\mega\hertz} carrier frequency with \SI{6.5}{\mega\hertz} sampling frequencies. The transmitted \gls{ki} and \gls{si} power levels were varied with \SI{2}{\deci\bel} steps, resulting in measurement grids that cover most practical received \glspl{sinr}. A reference timing generator was used that provides coarse knowledge about the positions of the \gls{ki} and \gls{si} waveforms in the received signal but does not reduce the carrier nor sampling frequency offsets nor lessen the challenge in estimating them.

The \gls{ki} was generated using a pseudo-random number generator with normal distribution and then limited to approximately \SI{5}{\mega\hertz} bandwidth. This approach can be straightforwardly adopted in practice to provide authorized receivers, and authorized receivers only, with the means to locally generate the \gls{ki} that is to be canceled based on a pre-shared secret seed for the pseudo-random number generator. The \gls{si} was an OFDM signal composed of 192 subcarriers with \SI{24}{\kilo\hertz} spacing, 32 of the subcarriers were used as pilots, and the OFDM symbols were padded with $\sfrac{1}{16}$-length cyclic prefixes. The data was modulated using either 4\nobreakdash-, 8\nobreakdash-, 16\nobreakdash-, 32\nobreakdash-, 64\nobreakdash-, 128\nobreakdash-, or 256-QAM and the pilots using BPSK. The OFDM symbols were concatenated into five frames of 5000 symbols each, whereas the first symbol of each frame was considered a training symbol known to the receiver. The \gls{vss-fo-lms} forgetting factors were set to $\lambda_\mathrm{e} = 0.99995$, $\lambda_\mathrm{R} = 0.9995$, $\lambda_\mathrm{y} = 0.999$, and $\lambda_\upepsilon = \lambda_\upeta = 0.99999$. $M=5$ taps were used to model the over-the-cable channels with $K=8$ maximum number of decision feedback iterations.

\begin{figure}[t!]
	\centerline{\includegraphics{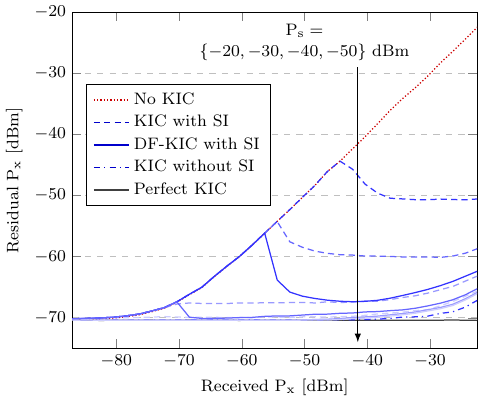}}
	\caption{Basic \gls{kic} performance depending on the received \gls{ki} and \gls{si} strength.}
	\label{fig:inr_analysis}
\end{figure}

\begin{figure}[tpb!]
	\centerline{\includegraphics{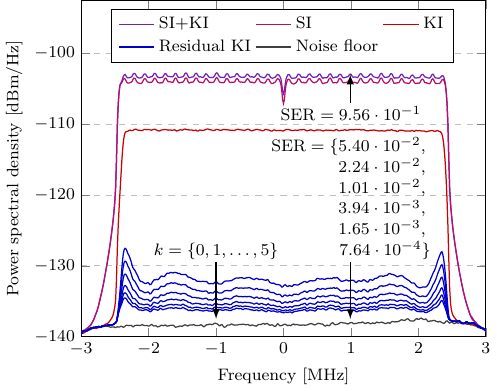}}
	\caption{Power spectral density of \gls{ki} throughout the \gls{df-kic} iterations.}
	\label{fig:psd_plot}
\end{figure}

\begin{figure*}[b!]
	\centerline{\includegraphics{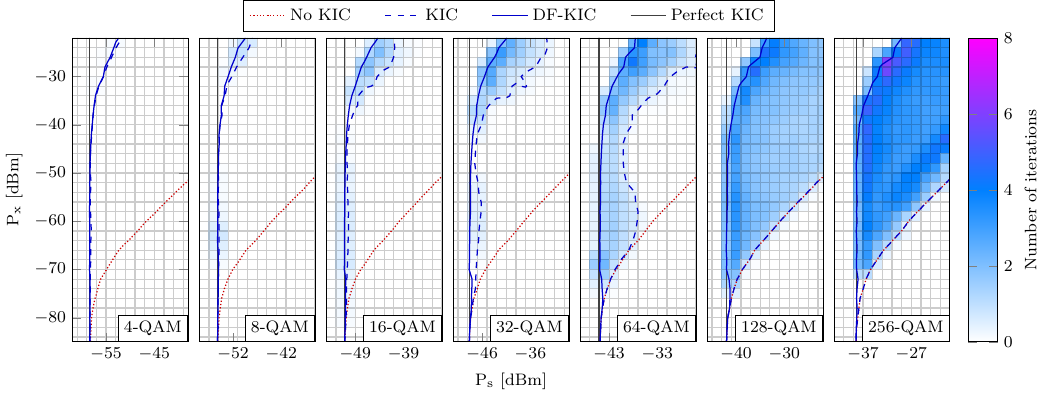}}
	\caption{Received \gls{sinr} requirements to reach $10^{-3}$ \gls{ser} depending on the receiver capabilities and \gls{si} modulation order. Also plotted are the number of decision feedback iterations required on average by the \gls{df-kic} to reach a \gls{ser} of $10^{-3}$.}
	\label{fig:ser_analysis}
\end{figure*}

\section{Measurement Results}
\label{sec:measurement_results}

Fig.~\ref{fig:inr_analysis} illustrates the \gls{si}'s impact on the base \gls{kic}. When no \gls{si} is received, the \gls{ki} is suppressed to the noise floor in most of the analyzed range. Only when the received \gls{ki} is very powerful does the cancellation become limited by the hardware impairments. However, when a \gls{si} is received, the \gls{ki} suppression is limited because of the interference from the \gls{si}. The residual \gls{ki} power level then depends on that of the received \gls{si}, e.g., the results show that when the received \gls{si} has a roughly \SI{50}{\deci\bel} \gls{snr}, the residual \gls{ki} after base \gls{kic} is at least \SI{20}{\deci\bel} above the noise floor, which, consequently, results in at most \SI{30}{\deci\bel} post-\gls{kic} \gls{sinr}. The residual \gls{ki} after \gls{df-kic}, however, is at most \SI{8}{\deci\bel} above the noise floor, which results in at most \SI{42}{\deci\bel} post-\gls{kic} \gls{sinr}.

The impact of decision feedback on \gls{kic} is further visualized in Fig.~\ref{fig:psd_plot}. In that measurement, the \gls{ki} and \gls{si} are received at \SI{-44}{\deci\belmilliwatt} and \SI{-38}{\deci\belmilliwatt} power levels, respectively, and the \gls{si} symbols are modulated using 256-QAM. As such, the \gls{ki} is powerful enough to completely prevent the \gls{si} from being demodulated and the \gls{si} is powerful enough to prevent the \gls{ki} from being completely suppressed. Through the decision feedback iterations, the \gls{kic} suppresses the residual \gls{ki} enough for the \gls{si} to be demodulated with a \gls{ser} below $10^{-3}$, which we here assume to be sufficiently low for regular post-\gls{kic} channel coding to handle the remaining errors.

Fig.~\ref{fig:ser_analysis} shows, depending on the receiver \gls{kic} capability and \gls{si} modulation order, the received \glspl{sinr} at which the $10^{-3}$ \gls{ser} can be reached and the number of decision feedback iterations that \gls{df-kic} takes on top of the base \gls{kic} to reach that target. The \glspl{sinr} are shown as contour lines and the number of required iterations are shown by the underlying heatmap. Since 4\nobreakdash- and 8-QAM have relatively low \gls{sinr} requirements, the post-\gls{kic} \gls{sinr} after applying only the base \gls{kic} once is essentially already sufficient to reach the targeted \gls{ser}. However, as the modulation order increases, the benefit of \gls{df-kic} over \gls{kic} becomes more and more evident, until 128\nobreakdash- and 256-QAM, when the base \gls{kic} alone is completely inadequate and the decision feedback iterations are indispensable in reaching the targeted \gls{ser}. The average number of iterations, considering only cases where \gls{df-kic} improves on \gls{kic}, is 0, 1, 1.4, 1.5, 1.6, 1.9, and 3.6 for 4\nobreakdash-, 8\nobreakdash-, 16\nobreakdash-, 32\nobreakdash-, 64\nobreakdash-, 128\nobreakdash-, and 256\nobreakdash-QAM, respectively.

\begin{figure}[tpb!]
	\centerline{\includegraphics{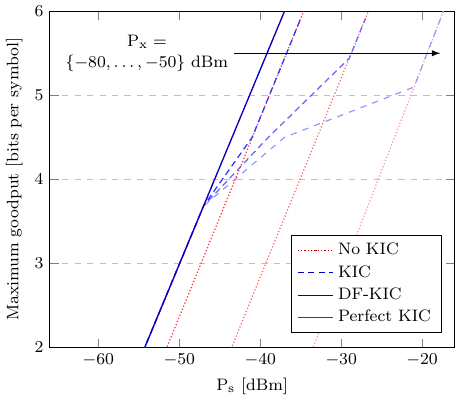}}
	\caption{Maximum achievable useful throughput.}
	\label{fig:max_bps_analysis}
\end{figure}

Finally, Fig.~\ref{fig:max_bps_analysis} demonstrates the impact of \gls{df-kic} on the communication system's useful throughput through
\begin{equation}
\text{goodput} = (1 - \text{SER}) \cdot \log_2(M^\textrm{QAM}) \cdot R,
\end{equation}
where $M^\textrm{QAM}$ is the modulation order and $R = \sfrac{3}{4}$ is the error correction coding rate. The results show that when the \gls{ki} is received weakly relative to the \gls{si}, the choice of \gls{kic} does not impact the achievable useful throughput (i.e., the resulting plot lines essentially overlap). However, at higher received \gls{ki} power levels, decision feedback iterations increase the number of useful bits that can be transmitted in a symbol by up to 35\%.

\section{Conclusion}
\label{sec:conclusion}

This letter introduced a \gls{df-kic} method that addresses the feasibility of \gls{kic} when the received \gls{ki} and \gls{si} are both powerful enough to interfere with the processing of one another. The proposed method iteratively cancels the received \gls{ki} and \gls{si} waveforms, assuming \textit{a priori} knowledge of the transmitted \gls{ki} and reconstructing the transmitted \gls{si} based on that received. Measurement results demonstrate that \gls{df-kic} outperforms basic \gls{kic} at many \glspl{sinr}, leading to lower post-\gls{kic} residual \gls{ki} and \gls{ser}. The improved performance comes at the expense of additional computational cost.

\ifCLASSOPTIONcaptionsoff
  \newpage
\fi

\IEEEtriggeratref{16}
\bibliographystyle{IEEEtran}
\bibliography{IEEEabrv,paper}

\end{document}